\documentclass[
    ,final       ]
  {aipproc}

\layoutstyle{6x9}

\usepackage{bm}

\begin{document}

\newcommand{\be}{\begin{equation}}
\newcommand{\ee}{\end{equation}}
\newcommand{\bea}{\begin{eqnarray}}
\newcommand{\eea}{\end{eqnarray}}
\newcommand{\bfk}{\bm{k}}
\newcommand{\pup}{p^\uparrow}
\newcommand{\pdown}{p^\downarrow}
\newcommand{\qup}{q^\uparrow}
\newcommand{\qdown}{q^\downarrow}
\newcommand{\bfp}{\bm{p}}
\newcommand{\nd}{\noindent}
\newcommand{\la}{\lambda}
\newcommand{\NP}[1]{{\it Nucl.\ Phys.}\ {\bf #1}}
\newcommand{\ZP}[1]{{\it Z.\ Phys.}\ {\bf #1}}
\newcommand{\PL}[1]{{\it Phys.\ Lett.}\ {\bf #1}}
\newcommand{\PR}[1]{{\it Phys.\ Rev.}\ {\bf #1}}
\newcommand{\PRL}[1]{{\it Phys.\ Rev.\ Lett.}\ {\bf #1}}
\newcommand{\MPL}[1]{{\it Mod.\ Phys.\ Lett.}\ {\bf #1}}
\newcommand{\SNP}[1]{{\it Sov.\ J.\ Nucl.\ Phys.}\ {\bf #1}}
\newcommand{\EPJ}[1]{{\it Eur.\ Phys.\ J.}\ {\bf #1}}
\newcommand{\IJMP}[1]{{\it Int.\ J.\ Mod.\ Phys.}\ {\bf #1}}

\vspace*{-1cm}

\title{Constraints on Gluon Sivers Distribution from RHIC Results\footnote{Talk
  delivered by U.~D'Alesio at the ``17th International Spin Physics
  Symposium'', SPIN2006, October 2-7, 2006, Kyoto, Japan.}}

\classification{12.38.Bx, 13.88.+e, 13.85.Ni, 14.70.Dj}

\keywords {single spin asymmetries, TMD distributions, Sivers effect}

\author{M. Anselmino}{
  address={Dipartimento di Fisica Teorica, Universit\`a di Torino and \\
INFN, Sezione di Torino, Via P. Giuria 1, I-10125 Torino, Italy}
}

\author{U. D'Alesio}{
  address={Dipartimento di Fisica, Universit\`a di Cagliari and \\
INFN, Sezione di Cagliari, C.P. 170, I-09042 Monserrato (CA),
Italy}
}

\author{S. Melis}{
  address={Dipartimento di Fisica, Universit\`a di Cagliari and \\
INFN, Sezione di Cagliari, C.P. 170, I-09042 Monserrato (CA),
Italy}
}

\author{F. Murgia}{
  address={Dipartimento di Fisica, Universit\`a di Cagliari and \\
INFN, Sezione di Cagliari, C.P. 170, I-09042 Monserrato (CA),
Italy}
}

\begin{abstract}
We consider the recent RHIC data on the transverse single spin asymmetry (SSA) $A_N$,
measured in $\pup p \to \pi^0 \, X$ processes at mid-rapidity by the PHENIX Collaboration.
We analyze this experimental information
within a hard scattering approach based on a generalized QCD factorization scheme, with
unintegrated, transverse momentum dependent (TMD), parton distribution and fragmentation
functions. In this kinematical region, only the gluon Sivers
effect could give a large contribution to $A_N$; its vanishing value
is thus used to give approximate upper limits on the gluon Sivers
function (GSF). Additional constraints from the Burkardt sum
rule for the Sivers distributions are also discussed.
\end{abstract}

\maketitle


\section{Formal approach}
\label{intr}

Transverse single spin asymmetries can originate, even with a short distance helicity
conserving pQCD dynamics, from spin-$\bfk_\perp$ correlations in the soft components of the
hadronic process $A^{\uparrow}B \to C\,X$. According to the hard scattering approach to
hadronic interactions developed in Refs.~\cite{fu,noi-1,noi-2}, based on the assumption of a
generalized QCD factorization scheme which involves unintegrated TMD parton distribution and
fragmentation functions, the general structure of the cross section for the polarized
hadronic process $(A,S_A) + (B,S_B) \to C + X$ reads (see Ref.~\cite{noi-2} for more
details)
\bea
\frac{E_C \, d\sigma^{(A,S_A) + (B,S_B) \to C + X}}
{d^{3} \bfp_C}  &=& \sum_{a,b,c,d, \{\la\}} \int
\frac{dx_a \, dx_b \, dz}{16 \pi^2 x_a x_b z^2  s} \;
d^2 \bfk_{\perp a} \, d^2 \bfk_{\perp b}\, d^3 \bfk_{\perp C}\,
\delta(\bm{k}_{\perp C} \cdot \hat{\bm{p}}_c)
\nonumber \\
 &\times& J(k_{\perp C})\,\rho_{\la^{\,}_a,
\la^{\prime}_a}^{a/A,S_A} \, \hat f_{a/A,S_A}(x_a,\bfk_{\perp a})
\> \rho_{\la^{\,}_b, \la^{\prime}_b}^{b/B,S_B} \,
\hat f_{b/B,S_B}(x_b,\bfk_{\perp b}) \nonumber\\
 &\times&
\hat M_{\la^{\,}_c, \la^{\,}_d; \la^{\,}_a, \la^{\,}_b} \,
\hat M^*_{\la^{\prime}_c, \la^{\,}_d; \la^{\prime}_a,
\la^{\prime}_b} \> \delta(\hat s + \hat t + \hat u) \> \hat
D^{\la^{\,}_C,\la^{\,}_C}_{\la^{\,}_c,\la^{\prime}_c}(z,\bfk_{\perp C})
\>, \label{gen1}
\eea
where all parton intrinsic motions are fully taken into account, both in the soft, non
perturbative components and in the hard, pQCD interactions.

The main features of Eq.~(\ref{gen1}) are the appearance of several spin and $\bfk_\perp$
dependent distribution and fragmentation functions and the non-collinear partonic
configuration which lead to many $\bfk_\perp$ dependent phases. In Ref.~\cite{noi-2} it was
explicitly shown that the only sizeable contributions to the transverse single spin asymmetry
$A_N(p^{\uparrow}p\to\pi\,X)$, in the kinematical region of large positive $x_F$ come from
the Sivers~\cite{siv} and, less importantly, from the Collins~\cite{col} mechanisms.
Moreover, while the quark contribution is totally dominant at large, positive $x_F$ values
(for polarized protons moving along the positive $Z$-axis), the gluon contribution may be
sizeable in the mid-rapidity and negative $x_F$ regions.

Data in the mid-rapidity region are available from the E704~\cite{e704y0} and
PHENIX~\cite{phe} experiments; the region of negative values of $x_F$ has been covered by the
STAR~\cite{star06} and BRAHMS experiments~\cite{brahms}. In these kinematical regions
$A_N(\pup p\to\pi\,X)$ is largely dominated by the Sivers effect~\cite{siv}, all other
contributions being almost vanishing, and Eq.~(\ref{gen1}) gives \cite{fu}:
\bea
\!\!\!\!\!\!\!\!\!\! && \frac{E_\pi \, d\sigma^\uparrow}{d^3\bfp_\pi} -
\frac{E_\pi \, d\sigma^\downarrow}{d^3\bfp_\pi} \simeq \sum_{a,b,c,d}
\int \frac{dx_a \, dx_b \, dz}{\pi \, x_a \, x_b \, z^2 \, s} \;
d^2\bfk_{\perp a} \, d^2\bfk_{\perp b} \, \,d^3\bfk_{\perp \pi} \,
\delta(\bfk_{\perp \pi}\cdot \hat{\bfp}_c) \, J(k_{\perp \pi})
\nonumber \\
\!\!\!\!\!\!\!\!\!\! &\times& \!\!\!\!\Delta \hat f_{a/\pup}(x_a, \bfk_{\perp a})
\> \hat f_{b/p}(x_b, k_{\perp b})\,\hat s^2 \>\frac{d\hat\sigma^{ab \to cd}}
{d\hat t}(x_a,x_b, \hat s, \hat t, \hat u) \> \delta(\hat s + \hat t +
\hat u) \> \hat D_{\pi/c}(z, k_{\perp \pi}) \,, \label{sivgen}
\eea
where
\be
\Delta \hat f_{a/\pup}\,(x_a, \bfk_{\perp a}) \equiv
\hat f_{a/\pup}\,(x_a, \bfk_{\perp a}) - \hat f_{a/p^\downarrow}\,
(x_a, \bfk_{\perp a})
\label{defsiv}
= \Delta^N \hat f_{a/\pup}\,(x_a, k_{\perp a}) \> \cos\phi_a \>.
\ee
$\Delta^N \hat f_{a/\pup}(x_a, k_{\perp a})$
[or $f_{1T}^\perp (x_a, k_{\perp a})]$ is referred to as the Sivers
distribution function of parton $a$ inside a transversely polarized (along the
$Y$-axis) proton (moving along the $Z$-axis).
$\phi_a$ is the azimuthal angle of the intrinsic transverse momentum
$\bfk_{\perp a}$ of parton $a$.

The azimuthal phase factor $\cos\phi_a$ appearing in the numerator of $A_N$,
Eqs.~(\ref{sivgen}) and (\ref{defsiv}), plays a crucial role and deserves a comment. The only
other term depending on $\phi_a$ in Eq.~(\ref{sivgen}) is the partonic cross section, in
particular via the corresponding Mandelstam variable $\hat{t}$. Therefore while at large
positive $x_F$ ($t$-channel dominated) the integration over $\phi_a$ does not necessarily
suppress $A_N$, for negative values of $x_F$ ($u$-channel dominated) one is roughly left with
the $d^2\bm{k}_{\perp a}\,\cos\phi_a$ integration alone, which cancels the potentially large
Sivers contribution. As a consequence, one cannot get significant information on the gluon
Sivers distribution from the recent STAR and BRAHMS data at negative values of $x_F$. (Notice
that at lower values of $\sqrt{s}$ the suppression induced by the $\cos\phi_a$ dependence
would be less drastic.)

The same arguments do not apply to inclusive hadronic processes at mid-rapidity and
moderately large $p_T$ values, for which data from PHENIX~\cite{phe} are already available,
for neutral pions and charged hadron production.  For these processes, the gluon contribution
is dominant and the Sivers effect can survive the phase integration. The possibility
of accessing the gluon Sivers function has been also investigated in Refs.~\cite{noi-3,b-v,sof}.

Indirect constraints on the GSF could also be obtained from a sum rule for the Sivers
distribution recently derived by Burkardt~\cite{burk}. The Burkardt Sum Rule (BSR) states
that the total (integrated over $x$ and $\bm{k}_\perp$) transverse momentum of all partons
(quarks, antiquarks and gluons) in a transversely polarized proton must be zero,
\begin{equation}
 \langle\bm{k}_{\perp}\rangle =
 \sum_a \;\langle\bm{k}_{\perp}\rangle_a =
 \int\!dx \int d^{\,2}\bm{k}_\perp \,\bm{k}_\perp
 \sum_a \Delta \hat{f}_{a/p^{\uparrow}}(x,\bm{k}_\perp) = 0 \, .
 \label{rule}
\end{equation}

In the following we shall simply check whether or not the proposed parameterizations of the
Sivers functions are compatible with the BSR.

\section{Phenomenology}
\label{ris}

We consider the PHENIX data \cite{phe} on $A_N$ for the $\pup p \to \pi^0 \, X$ process at
RHIC, at $\sqrt s$ = 200 GeV, with $p_T$ ranging from 1.0 to 5.0 GeV/$c$ and mid-rapidity
values, $|\eta| < 0.35$. In this kinematical regime, at the lowest $p_T$ values,
$x_{a}^{\rm{min}}$ can be as small as 0.005. Therefore, partonic channels involving a gluon
in the transversely polarized initial proton dominate over those involving a quark. This
gluon dominance, together with the (almost) vanishing of all possible contributions to $A_N$
other than the Sivers effect, allows to put upper bounds on the GSF. Due to possible mixing
with quark initiated contributions, the same is not true for the E704 data~\cite{e704y0} at
lower energies and comparable rapidity and $p_T$ ranges.

In Ref.~\cite{fu} we have shown that reasonable fits to the SSA for the $p^\uparrow p\to \pi
X$ process at large positive $x_F$ can be obtained by using valence-like Sivers functions for
$u$ and $d$ quarks, which turn out to have opposite signs. The use of valence-like $u$ and
$d$ Sivers functions alone predicts an almost vanishing SSA in the mid-rapidity region,
compatible with available data (see Fig.~1(left), for the PHENIX results). Their
parameterizations~\cite{fu,noi-4} are also compatible with the BSR, Eq.~(\ref{rule}).
Although a large gluon Sivers function would not modify the analysis of the SIDIS, E704 and
STAR data at large positive $x_F$, it would strongly affect the description of the
mid-rapidity PHENIX data.

In what follows (see Ref.~\cite{noi-gsf} for details)  
we therefore try to understand what is the maximum value of $|\Delta^N \hat
f_{g/\pup}\,(x,k_{\perp})|/2\hat f_{g/p}\,(x, k_{\perp})$ allowed by the PHENIX data; our
results are summarized in Fig.~1 (left panel for the SSA and right panel for the GSF):\\
\noindent $\bullet$ The thin, solid line in Fig.~1(left), results from computing
$A_N$ using only the valence-like $u$ and $d$ Sivers functions of Ref.~\cite{fu}.\\
$\bullet$ The dot-dashed curve in Fig.~1(left) has been obtained by saturating (in magnitude)
the GSF to the natural positivity bound [see Eq.~(\ref{defsiv})]
\begin{equation}
\Delta^N \hat f_{g/\pup}\,(x,k_{\perp}) = - 2\hat f_{g/p}\,(x, k_{\perp}) \>. \label{gsfmax}
\end{equation}
The sea-quark Sivers functions are again assumed to vanish. This choice leads to a SSA
definitely in contradiction with data and to a strong violation of the BSR. \\
$\bullet$ The thick,  solid curve in Fig.~1(left) has been obtained still assuming that there
is no sea-quark Sivers contribution, and looking for a parameterization of $\Delta^N \hat
f_{g/\pup}$ yielding values of $A_N$ falling, approximately, within one-sigma deviation below
the lowest $p_T$ data. The corresponding $x$-dependent part of the GSF, normalized to its
positivity bound, $|\Delta^N f_{g/\pup}(x)|/ 2 f_{g/p}(x)$, is shown as the solid curve in
Fig.~1(right). It leads, within the $x$ range covered by the data, to a strong violation of
the BSR.\\
$\bullet$ We finally consider the inclusion of all sea-quarks ($u_s$, $\bar{u}$, $d_s$,
$\bar{d}$, $s$, $\bar{s}$) contributions by using a non-vanishing positive Sivers function
which saturates the positivity bound [that is, $\Delta^N \hat f_{q_s/\pup}\,(x, k_{\perp})
\equiv 2\hat
  f_{q_s/p}\,(x, k_{\perp})$].
These contributions could then cancel the negative contribution to $A_N$ of a possibly large
GSF. We then look for the largest negative GSF which, together with a positive maximized
sea-quark contribution, leads again to the SSA represented by the thick, solid line of
Fig.~1(left). This curve results now as the sum of the (maximized) sea and valence quark
contribution (dotted curve) and that of the new GSF (dashed curve), which is plotted as the
dashed curve in Fig.~1(right). It is the largest (overmaximized) gluon Sivers function
compatible with PHENIX data. Within the $x$ range covered by the data, the (over)maximized
sea-quark Sivers distributions give a positive contribution which strongly suppresses the
negative contribution of the GSF, so that in this scenario the BSR is satisfied within a 10\%
level of accuracy.

Summarizing, our analysis shows that the PHENIX data on $A_N(p^{\uparrow}p\to\pi^{0}\,X)$
allow to put significant quantitative bounds on the magnitude of the GSF. Similar conclusions
have been recently reached by studying the Sivers effect in SIDIS off a deuteron
target~\cite{bg}.

\begin{center}
\begin{figure}[t]
\includegraphics[height=.3\textheight,angle=-90]{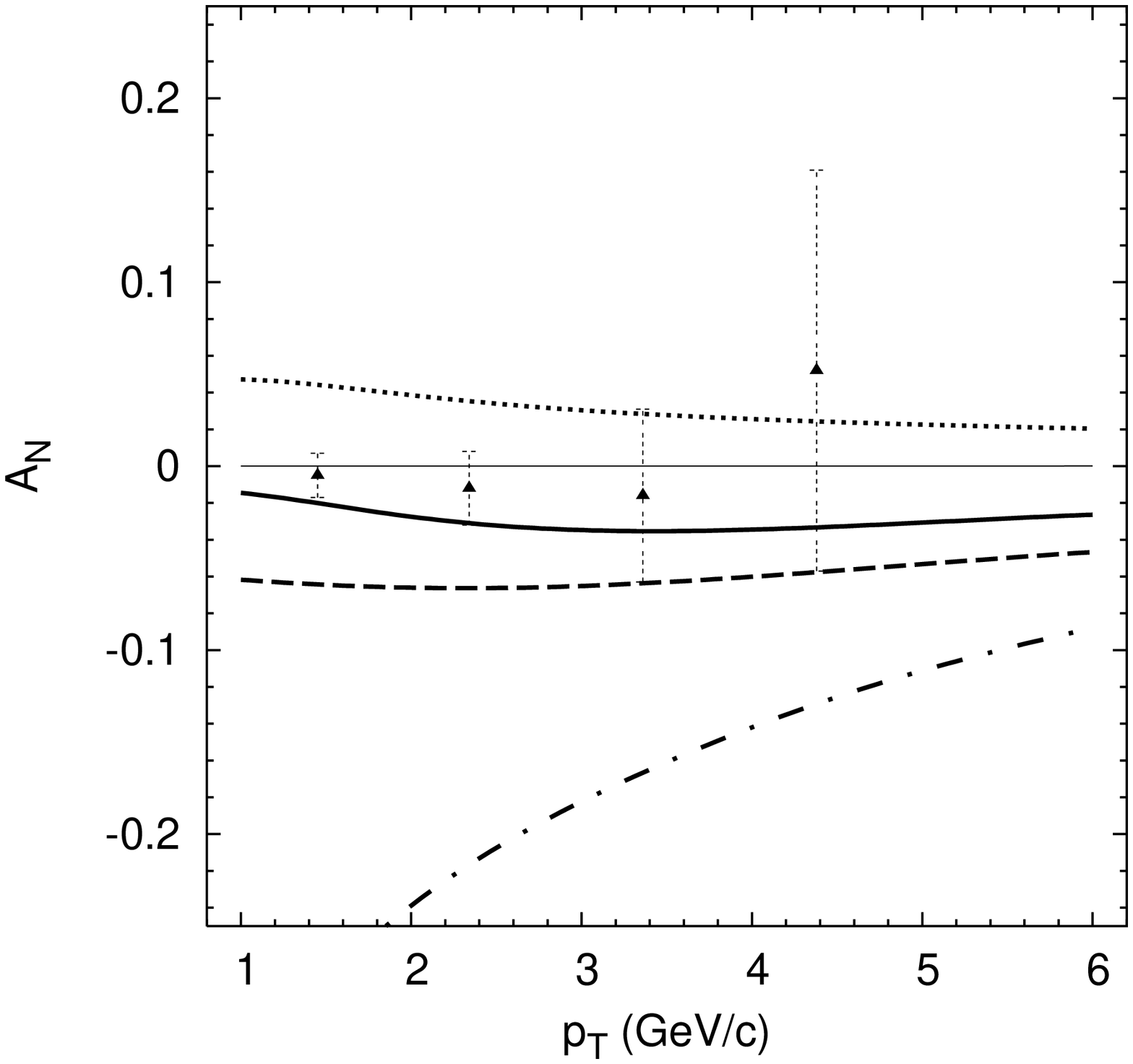}
\includegraphics[height=.3\textheight,angle=-90]{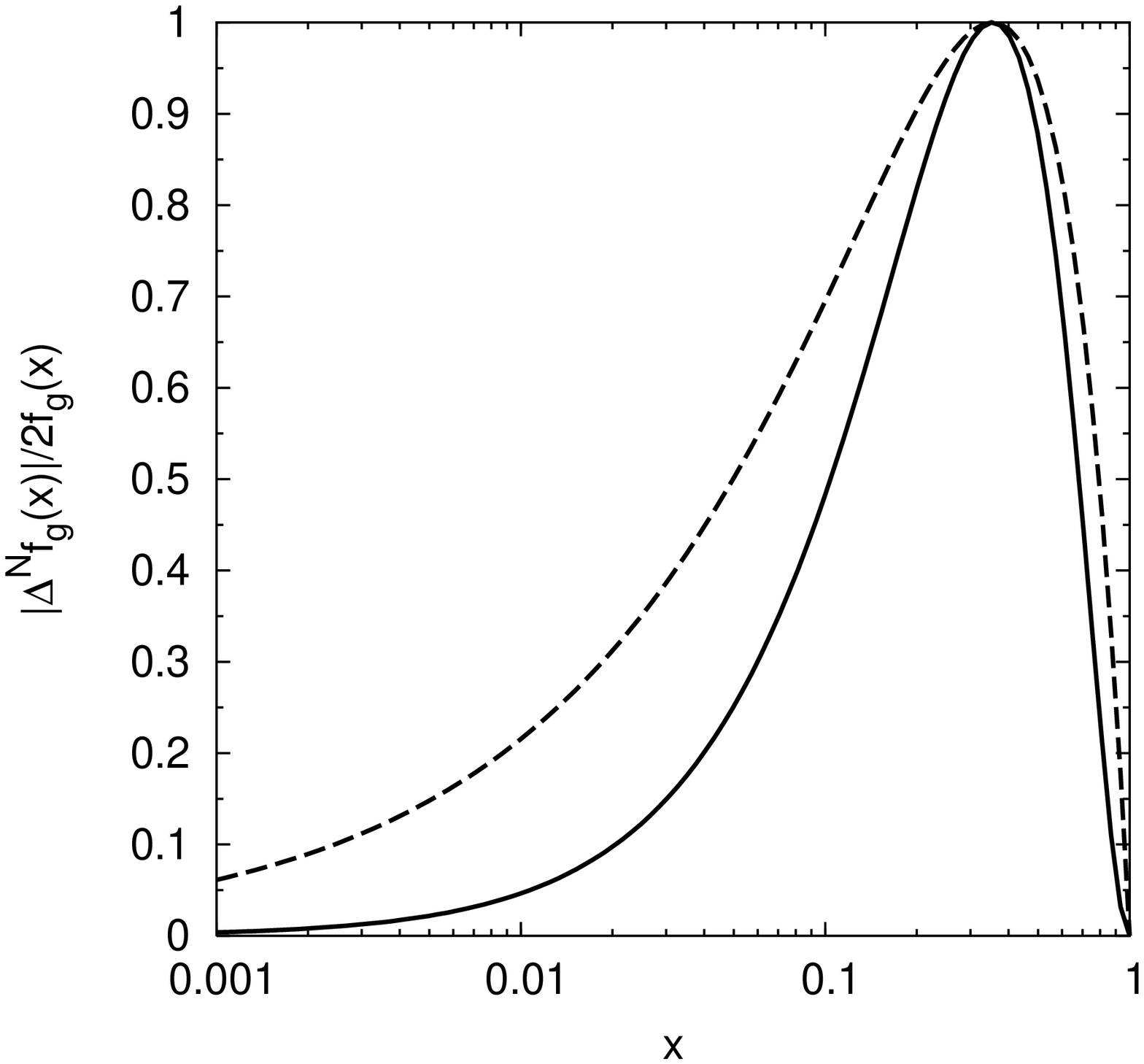}
\caption{Left: the computed SSA, $A_N$, compared
with PHENIX data \cite{phe}, with different choices for the
gluon and sea-quark Sivers functions.
Right: the value of the normalized GSF, $|\Delta^N f_{g/\pup}(x)|
/2 f_{g/p}(x)$.
}
\end{figure}
\end{center}

\bibliographystyle{aipproc}   

\end{document}